\documentclass[conference]{IEEEtran}
\IEEEoverridecommandlockouts
\usepackage{cite}
\usepackage{amsmath,amssymb,amsfonts}
\usepackage{algorithmic}
\usepackage{graphicx}
\usepackage{textcomp}
\usepackage{xcolor}
\usepackage[most]{tcolorbox}
\usepackage{tabularx}
\usepackage{enumitem}  
\usepackage{xcolor}    
\usepackage{booktabs}
\usepackage{pifont}
\usepackage{multirow}
\usepackage{makecell}
\usepackage{amssymb}
\usepackage{url}

\def\BibTeX{{\rm B\kern-.05em{\sc i\kern-.025em b}\kern-.08em
    T\kern-.1667em\lower.7ex\hbox{E}\kern-.125emX}}
\begin{document}

\title{Multi-CoLoR:  Context-Aware Localization and Reasoning across Multi-Language Codebases }

\author{
\IEEEauthorblockN{ \small
Indira Vats\IEEEauthorrefmark{1}\IEEEauthorrefmark{2},
Sanjukta De\IEEEauthorrefmark{1},
Subhayan Roy\IEEEauthorrefmark{1},
Saurabh Bodhe\IEEEauthorrefmark{1},
Lejin Varghese\IEEEauthorrefmark{1},
Max Kiehn\IEEEauthorrefmark{1},
Yonas Bedasso\IEEEauthorrefmark{1},
Marsha Chechik\IEEEauthorrefmark{2}
}
\IEEEauthorblockA{\footnotesize
\IEEEauthorrefmark{1}Advanced Micro Devices, Inc. (AMD)\quad
\IEEEauthorrefmark{2}University of Toronto\\
Emails: \{indiravats, chechik\}@cs.toronto.edu,
\{Sanjukta.De, Subhayan.Roy, Saurabh.Bodhe, Lejin.Varghese, max.kiehn, Yonas.Bedasso\}@amd.com
}
}

\maketitle

\begin{abstract}
Large language models demonstrate strong capabilities in code generation but struggle to navigate complex, multi-language repositories to locate relevant code. Effective code localization requires understanding both organizational context (e.g., historical issue–fix patterns) and structural relationships within heterogeneous codebases. Existing methods either (i) focus narrowly on single-language benchmarks, (ii) retrieve code across languages via shallow textual similarity, or (iii) assume no prior context. We present \texttt{Multi-CoLoR}, a framework for \underline{Co}ntext-aware \underline{Lo}calization and \underline{R}easoning across \underline{Multi}-Language codebases, which integrates organizational knowledge retrieval with graph-based reasoning to traverse complex software ecosystems. Multi-CoLoR operates in two stages: (i) a similar issue context (SIC) module retrieves semantically and organizationally related historical issues to prune the search space, and (ii) a code graph traversal agent (an extended version of LocAgent, a state-of-the-art localization framework) performs structural reasoning within C++ and QML codebases. Evaluations on a real-world enterprise dataset show that incorporating SIC reduces the search space and improves localization accuracy, and graph-based reasoning generalizes effectively beyond Python-only repositories. Combined, Multi-CoLoR improves Acc@5 over both lexical and graph-based baselines while reducing tool calls on an AMD codebase. 
\end{abstract}

\begin{IEEEkeywords}
agentic reasoning, code localization, graph-based reasoning, LLM, organization memory, semantic search
\end{IEEEkeywords}

\vspace{-0.1in}
\section{Introduction}
\label{sec:introduction}
Recent advances in large language models (LLMs) have significantly improved automated code generation~\cite{10.1145/3747588}. However, these models exhibit notable limitations in accurately identifying relevant code regions within complex software systems. This task, known as code localization, involves determining the precise location for applying bug fixes. It plays a pivotal role in automated program repair (APR) as localization accuracy directly influences downstream repair success~\cite{liu2019you}. Given the inherent complexity of localization due to code interdependencies and semantic ambiguity, recent work~\cite{hossain2024deep} has treated it as a standalone phase, decoupled from repair to enhance modularity and effectiveness.

Effective localization demands rich \textit{contextual} information, including historical knowledge of past issues and resolutions~\cite{8625380}~\cite{motwani2023better}, and a deep understanding of project structure and cross-file semantics~\cite{chen-etal-2025-locagent}. However, such information often exceeds the context window of LLMs, limiting their ability to reason effectively. While enriched context via bug reports or retrieval-augmented (RAG) methods~\cite{parvez-etal-2021-retrieval-augmented}~\cite{wang2023rap} have improved repair pipelines, multiple studies identify localization as a persistent bottleneck~\cite{chen-etal-2025-locagent}~\cite{wang2025improving}, with evidence suggesting that context injection alone is insufficient  \cite{yu2025orcalocallmagentframework}.

Traditional LLM-based systems often rely on single-pass inference, limiting their ability to reason across complex, multi-language codebases. To address these limitations, recent work ~\cite{singh2025agenticretrievalaugmentedgenerationsurvey} has introduced \textit{agentic reasoning}, a paradigm where LLMs act as autonomous agents that plan actions, retrieve relevant context and observe intermediate outcomes to iteratively refine their strategies. This multi-step, tool-augmented reasoning enables deeper understanding and more accurate localization. 

In enterprise settings, code localization operates at fundamentally greater complexity compared to typical benchmark datasets (single-language, self-contained projects).  Real-world enterprise repositories span heterogeneous ecosystems, integrating multiple programming languages and extensive inter-module dependencies. Most existing approaches focus narrowly on single-language environments (typically Python) and fail to generalize~\cite{zan2025multi}. For instance, the SWE-bench dataset~\cite{ICLR2024_edac78c3} contains only Python issue–fix pairs, limiting its applicability to diverse enterprise software ecosystems. This monolingual focus creates a critical gap: real-world industrial codebases often combine languages (e.g., C++ backends, Python bindings, and QML frontends), yet current methods struggle to reason across such multi-language codebases.  

\begin{figure*} 
    \centering
    \includegraphics[width=0.9\textwidth, height=8cm]{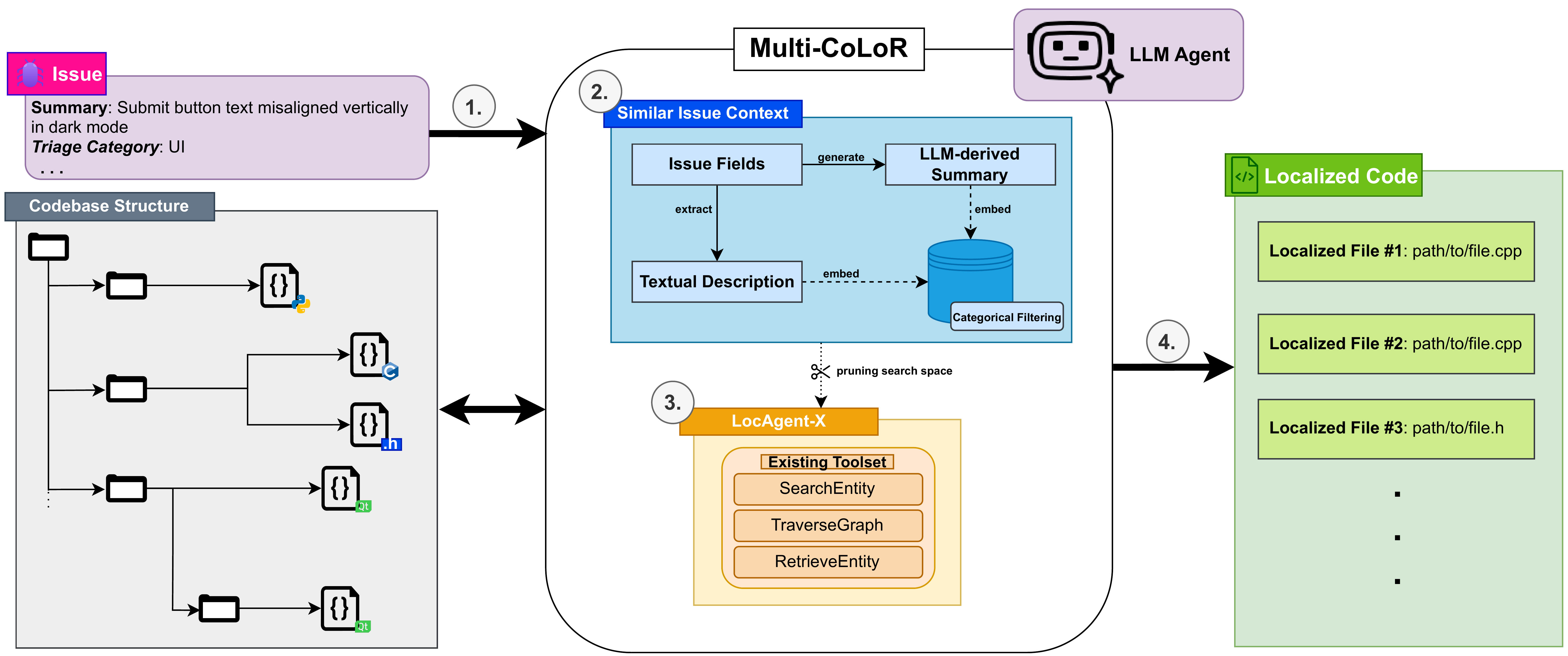}
    \caption{Multi-CoLoR pipeline for context-aware code localization and reasoning in multi-language repositories. \ding{172} A new issue is ingested. \ding{173} The similar issue module retrieves top-\textit{k} similar historical issues and artifacts (summaries, components, file paths), producing cues that condition the search. \ding{174} LocAgent-X parses the repository and builds a Unified Dependency Graph from the codebase structure (across QML/C++/Python) to localize the file. \ding{175} The system returns ranked fix locations.}
    \label{fig:architecture}
    \vspace{-0.1in}
\end{figure*}

Recent work~\cite{zan2025multi} confirms that even state-of-the-art code agents~\cite{yang2024swe, xia2024agentless, wang2024openhands}  degrade significantly on non-Python tasks, underscoring the challenges posed by cross-language tooling and semantic differences. Moreover, industrial datasets exhibit weaker overlap between bug descriptions and code compared to academic benchmarks~\cite{11121734}, necessitating more sophisticated fault localization techniques that can actually \textit{reason} on the codebase. While some methods are language-agnostic~\cite{10.1145/2786805.2786855}, their simplistic architecture fails to capture the deeper semantic and structural signals required for robust localization.

Building on these insights, we posit that effective localization in industrial settings requires \textit{context-aware reasoning} that transcends textual similarity and isolated code retrieval. Real-world repositories span  loosely coupled components across multiple languages, embedded within complex organizational contexts, where crucial cues lie in issue reports, commit histories, and recurring fix patterns.

To address these challenges, we propose \textbf{\texttt{Multi-CoLoR}}, a framework for \underline{Co}ntext-aware \underline{Lo}calization and \underline{R}easoning across large \underline{Multi}-language codebases. \texttt{Multi-CoLoR} integrates organizational knowledge with graph-based reasoning. As shown in Figure~\ref{fig:architecture}, our approach operates in two stages: (i) a Similar Issue Context (SIC) module retrieves semantically and organizationally related historical issues to prune the search space; and (ii) an extended version of LocAgent~\cite{chen-etal-2025-locagent} performs graph-based reasoning over a unified dependency graph constructed from the codebase to localize candidate files.

This study is guided by three research questions:

\begin{description}
    \item[RQ1:] Does incorporating similar historical issue knowledge as contextual input improve localization accuracy and effectively prune the search space?
    
    \item[RQ2:] Can graph-based code-reasoning techniques generalize beyond Python and handle industrial datasets with sparse lexical overlap between issue descriptions and code?
    
    \item[RQ3:] What is the combined impact of organizational knowledge and  graph-based reasoning on localization performance in large, heterogeneous repositories?
\end{description}

We focus on a concrete instantiation of Multi-CoLoR in an industrial setting at AMD. We evaluate Multi-CoLoR on a large enterprise codebase with components in C++ and QML. On a dataset of 2{,}563 issues with verified ground-truth localizations, the SIC module achieves high component- and directory-level match rates and substantially prunes the search space even when issue descriptions are sparse. On a 140-issue C++/QML subset,  Multi-CoLoR attains the highest Acc@5 while reducing the number of tool calls compared to both lexical and graph-based baselines.

In summary, this work makes the following contributions:

\begin{enumerate}
    \item We propose Multi-CoLoR, a unified context-aware framework that integrates organizational knowledge retrieval with agentic reasoning over code graphs for multi-language code localization.

    \item We empirically validate Multi-CoLoR on industrial data from AMD, quantifying the benefits of SIC-driven contextual augmentation in code fix localization.
    
    \item We extend LocAgent to enable graph-based localization beyond Python with support for mixed C++/QML repositories via a unified dependency graph.
        
    \item We distill practical design insights for building agentic code-localization systems in large industrial repositories, including guidance on modular tool composition and deployment trade-offs.
\end{enumerate}

The remainder of this paper is organized as follows: Sec.~\ref{sec:related} reviews related work on code localization techniques; Sec.~\ref{sec:methodology} 
details our methodology; Sec.~\ref{sec:similar-issue} and 
\ref{sec:extended-locagent} explain the two main components of our framework; Sec.~\ref{sec:evaluation} 
presents experimental setup, results and threats to validity; Sec.~\ref{sec:discussion} discusses  lessons learned and Sec.~\ref{sec:conclusion} 
concludes with future directions.

\section{Related Work}
\label{sec:related}

{\bf Non-agent LLM methods}
Existing non-agent approaches adopt simplified \textit{retrieve-and-edit} pipelines, coupling repository-aware retrieval with direct patch generation. \emph{Agentless}~\cite{xia2024agentless} eschews tool-using agents in favor of a simple three-stage process: localization, repair, and patch validation, demonstrating competitive results on SWE-bench without complex planning. Repository-level completion methods such as \emph{RepoCoder}~\cite{zhang-etal-2023-repocoder} use a similarity-based retriever with iterative retrieval–generation to surface cross-file context before producing completions, while \emph{RepoFuse}~\cite{liang2024repofuserepositorylevelcodecompletion} fuses analogy and rationale contexts and compresses them via rank-truncated generation to fit prompt budgets with low latency. These approaches reduce the overhead of multi-step reasoning but rely heavily on embedding retrieval quality and lack agentic exploration of large repositories, limiting effectiveness when issue descriptions are sparse.

{\bf Agent-based methods}
Agentic systems tackle repository-level issue resolution via iterative tool use and planning. \emph{SWE-Agent}~\cite{yang2024swe} formalizes an Agent–Computer Interface (file viewers, editors, search), substantially outperforming simple retrieval baselines albeit at higher cost. \emph{OpenHands}~\cite{wang2024openhands} (formerly \emph{OpenDevin}) provides a community-driven platform for agent development with an extensible skill library and optional multi-agent delegation; its generalist CodeActAgent attains competitive results across task categories when compared to specialized agents. To enhance exploration using backtracking and self-improvement, \emph{SWE-Search}~\cite{DBLP:conf/iclr/AntoniadesOZXGW25} integrates Monte Carlo Tree Search (MCTS) with Value and Discriminator agents for feedback and collaboration, improving over standard linear, sequential open-source agents. Despite improved end-to-end success, these systems treat localization as an \textit{emergent} byproduct of search-read-edit loops rather than a dedicated, structure-aware task. This coarse-grained approach limits precision in complex, multi-language repositories.

{\bf Graph-based agentic methods}
A complementary line builds explicit repository graphs to guide retrieval and reasoning. We use graph-based reasoning to refer to methods that construct an explicit code graph and expose graph-indexed tools so that the agent can traverse and query this structure during localization. 
\emph{CodexGraph}~\cite{liu2024codexgraphbridginglargelanguage} exposes a static code graph through a graph-database interface so the agent can issue structural queries for precise, multi-hop navigation to retrieve relevant code fragments or code structures. \emph{RepoGraph}~\cite{DBLP:conf/iclr/Ouyang0MX0J00025} is a plug-in module that parses a repository to create a line-level dependency graph and allows relevant subgraph retrieval \textit{(ego-graph retrieval)} in procedural pipelines and agents, yielding consistent gains on integration with both. \emph{OrcaLoca}~\cite{yu2025orcalocallmagentframework} similarly builds a \emph{CodeGraph} over files/classes/methods/functions with containment and reference edges and pairs it with a graph-backed search API, adding priority scheduling, decomposition, and context pruning to focus localization. \emph{LingmaAgent}~\cite{10.1145/3696630.3728549} (formerly \emph{RepoUnderstander}) builds a top–down repository structure tree, expands it into a function-level reference/call graph, and uses MCTS-guided exploration and summarization to condense repo-level context for localization and repair. \emph{LocAgent}~\cite{chen-etal-2025-locagent} is a state-of-the-art method that constructs a directed code graph and equips agents with graph-indexed tools for multi-hop traversal, reporting strong file-level localization. Collectively, these methods demonstrate the value of structure-aware reasoning; however, most evaluations are Python-focused and do not leverage enterprise history. Multi-CoLoR builds directly on LocAgent’s heterogeneous graph paradigm, extending it to multi-language settings and coupling it with similar-issue retrieval to inject organizational context into graph-guided localization.

{\bf Memory-enhanced methods}
Recent work augments LLM agents with structured context and memory to enable efficient reasoning. \emph{SWE-Exp}~\cite{chen2025swe} introduces an \textit{experience bank} that distills reusable patterns from prior repair trajectories (successful and failed), improving downstream resolution by experience-driven exploration rather than exploration from scratch. Building on graph agents, \emph{RepoMem}~\cite{wang2025improving} expands LocAgent's toolset by equipping it with memory lookup tools of commit history and key-file summaries, addressing the limitation that agents otherwise approach each issue independently. These methods highlight a shift toward continuity of reasoning across tasks. Our approach is complementary: in industrial settings where issue descriptions contain few code terms, we leverage \textit{organizational memory} via similar-issue retrieval and then reason over a \textit{multi-language} code graph.

{\bf Industrial prototypes}
Google’s \emph{Passerine}~\cite{11121734} provides an industrial baseline for agentic APR in a multilingual monorepo: a minimalist, ReAct-style agent integrated with internal code search and build/test tooling to operate across Java, C++, TypeScript, Kotlin, and Python. Evaluation on an internal dataset of 178 production issues (78 human-reported, 100 machine-reported) reveals that agent performance varies sharply between machine-generated and human-reported bugs. The authors attribute the difficulty of industrial bugs to distributional differences from benchmark datasets (language diversity, patch size/spread) and explicitly note that Passerine is purposefully simple, not explicitly tasked with localization, and does not exploit rich development histories (e.g., similar past bugs/patches) that could refine fault localization. Alibaba’s \emph{LingmaAgent}~\cite{10.1145/3696630.3728549} achieves automatic resolution of $16.9\%$ on a multilingual in-house dataset (Java, TypeScript, JavaScript), rising to $43.3\%$ with developer intervention~\cite{10.1145/3696630.3728549}, underscoring the added value of human-specific context that current agents do not exploit. Compared to Passerine’s intentionally minimalist design, LingmaAgent introduces explicit structure modeling, but its reference relations are restricted to the \emph{function} level to control graph complexity, and it does not leverage organizational history (e.g., similar past issues/patches) as a first-class signal. Multi-CoLoR addresses these gaps by coupling similar-issue retrieval with a richer, heterogeneous code graph that supports multi-language reasoning to strengthen localization in heterogeneous repositories.

{\bf Benchmarks, datasets, and language coverage}
\emph{SWE-bench}~\cite{jimenez2024swebench} is the de facto benchmark for repository-level issue resolving, pairing real GitHub issues with a reproducible, Dockerized evaluation harness. Derivatives broaden access and reliability: \emph{SWE-bench Lite} curates a 300-task subset for faster iteration, and \emph{SWE-bench Verified} provides a human-validated split for more trustworthy scoring. Recent analyses caution that leaderboard gains may be conflated with memorization and benchmark-specific contamination, undermining claims of proficiency~\cite{liang2025the}. To address the Python-only scope, \emph{SWE-bench Multilingual}~\cite{yang2025swesmith} releases 300 tasks across 9 languages with baseline results substantially lower than on Python-only splits, emphasizing task difficulty beyond Python. In addition, \emph{Multi-SWE-bench}~\cite{zan2025multi} introduces a multilingual suite spanning seven languages (Java, TypeScript/JavaScript, Go, Rust, C, C++), with 1{,}632 expert-annotated instances. Since SWE-bench primarily targets end-to-end repair with localization as an implicit step, \emph{Loc-Bench}~\cite{chen-etal-2025-locagent} focuses explicitly on \textit{localization}: it collects up-to-date Python issues to reduce pretraining bias and broadens task types (bugs, features, security, performance) for more comprehensive localization evaluation. These resources reveal persistent gaps: most public datasets remain Python-centric, and expose substantial performance drops outside Python. These benchmark datasets are also not reflective of the complexities of real-world industrial datasets.

\section{Methodology}
\label{sec:methodology}
We design a localization pipeline tailored to the constraints of industrial codebases: incomplete issue descriptions, heterogeneous repositories, and the need for organization-aware search. Our approach is guided by two key observations from enterprise development at AMD:  organizational signals create predictable patterns between issue types and code locations that pure semantic matching cannot capture, and  industrial repositories require heterogeneous reasoning capabilities absent from existing single-language localization frameworks.

A key design decision is to leverage the inherent organizational structure within enterprise development environments. Industrial software teams often exhibit specialized ownership patterns, where specific teams consistently handle particular types of issues. Issue tracking systems (e.g., Jira) capture this structure through metadata fields such as triage classifications and program names, which can be exploited as structured filters for narrowing candidate issues. Additionally, large industrial codebases exhibit hierarchical organization through nested directory structures that reflect architectural boundaries. While similar issues may not always require fixes in identical files, they consistently cluster within related subdirectories or components. By leveraging this to bound the search region, we constrain the search space from thousands of files to manageable subsets. Within these bounded regions, a reasoning-based localizer performs precise file-level identification.

For multi-language support, we extend the original Python-only LocAgent to build a single Unified Dependency Graph (UDG) spanning C++ and QML (and preserving existing Python support). The UDG captures intra-language relations (e.g., \texttt{IMPORTS}, \texttt{INHERITS}) and shared structural relations (e.g., \texttt{CONTAINS}).  We refer to the multi-language extension of LocAgent as LocAgent-X (LocAgent-e\underline{X}tended) to distinguish it from the original Python-only framework.

Multi-CoLoR operationalizes these insights though a sequential design. First, the SIC module retrieves top-\(k\) historical issues to produce cues that \textit{prune} the search space. Second, LocAgent-X performs precise localization within the pruned region using graph-based reasoning.

The integrated workflow is illustrated in Figure~\ref{fig:architecture}. Sec.~\ref{sec:similar-issue} and~\ref{sec:extended-locagent} detail SIC and LocAgent-X, respectively.

\section{Similar Issue Context}

\label{sec:similar-issue}
\subsection{Motivation}
Issue reports are semi-structured artifacts that combine free-text with ordered and categorical metadata. At AMD, the issue tracker is the primary channel for reporting defects and regressions. Each issue typically includes a natural-language summary and reproduction steps, alongside structured labels such as \textsc{Severity}, \textsc{Priority}, and \textsc{Triage Category}.

However, real-world issues are noisy and incomplete: fields like \textsc{program name} and \textsc{priority} are consistently populated, while others (e.g., \textsc{root cause}, \textsc{feature summary}) appear sporadically. This variability complicates similarity computation and downstream localization. Figure~\ref{fig:synthetic_issue_darkmode} shows a representative example.

\begin{figure}[t]
\centering
\begin{tcolorbox}[
  enhanced, colback=gray!5, colframe=black!30,
  boxrule=0.4pt, sharp corners,
  title=\small\bfseries TCKT-EXAMPLE-0451 \,·\, Defect \,(P2)]
\small
\begin{tabularx}{0.92\linewidth}{@{}l X@{}}
\textbf{Program} & PRODUCT\_Y \\
\textbf{Found In} & BRANCH\_REL\_Y \\
\textbf{Triage Category} & UI \\
\textbf{Created} & 2025-10-29 09{:}47 \\
\textbf{Assignee} & (Unassigned) \\
\end{tabularx}

\vspace{0.6ex}\hrule height 0.4pt \vspace{0.6ex}

\textbf{Title} \quad Submit button text misaligned vertically in dark mode.

\vspace{0.5ex}

\textbf{Description} \quad In dark mode, the “Submit” button label appears slightly lower than center, causing visual imbalance. Issue does not occur in light mode.

\vspace{0.5ex}

\textbf{Root Cause} $\quad \langle \mbox{\emph{missing}}\rangle$

\vspace{0.5ex}

\textbf{Feature Summary} \quad Button rendering and theme adaptation logic.

\vspace{0.6ex}\hrule height 0.4pt \vspace{0.6ex}

\textbf{Steps to Reproduce}
\begin{enumerate}[leftmargin=2.0em,itemsep=0.1ex,topsep=0.3ex]
  \item Launch PRODUCT\_Y.
  \item Navigate: Settings $\rightarrow$ Appearance $\rightarrow$ Enable Dark Mode.
  \item Go to: Form Wizard $\rightarrow$ Final Step.
  \item Observe “Submit” button label alignment.
\end{enumerate}

\textbf{Expected} \quad Button label should be vertically centered in both themes.

\textbf{Actual} \quad Label appears lower in dark mode, breaking visual symmetry.
\end{tcolorbox}
\vspace{-0.1in}
\caption{Sample issue illustrating a UI defect.}
\label{fig:synthetic_issue_darkmode}
\vspace{-0.1in}
\end{figure}

\vspace{-0.1in}
\subsection{Background: Duplicate Bug Report Detection (DBRD)}

The design of \textit{SIC} builds on the DBRD literature.

\paragraph{Classical retrieval} 
An early approach that combines text with categorical/ordinal fields. REP \cite{6100061} uses BM25F$_{\text{ext}}$ for free text and simple matching for metadata, REP achieved strong retrieval performance with minimal computational cost.

\paragraph{Deep learning methods} 
Subsequent studies adopted independent embeddings (siamese triplet, siamese pair \cite{8094414},  DBR-CNN \cite{8719497}, DWEN \cite{8449496}) and attention-based architectures \cite{poddar-etal-2019-train}, capturing semantic similarity across free-text fields. SABD \cite{10148717} introduced a two-branch cross-attention model jointly modeled text and metadata, improving accuracy but increasing inference latency due to limited pre-caching.

\vspace{-0.01in}
\paragraph{Hybrid LLM approaches} 
CUPID \cite{zhang2023cupid} demonstrated a hybrid approach, employing an LLM to extract concise, clean keywords from noisy text before ranking them with a classical retriever (REP). This approach preserved interpretability and speed while leveraging LLM reasoning to handle inconsistent field population.

\begin{table}[t]
\centering
\caption{Comparison of DBRD Methods.}
\resizebox{\columnwidth}{!}{
\begin{tabular}{|l|l|l|l|}
\hline
\textbf{Method} & \textbf{Fields Used} & \textbf{Strengths} & \textbf{Weaknesses} \\ \hline
REP & Text + Categorical + Ordinal & Fast, low resource & Limited semantics \\ \hline
SABD & Text + Categorical & Cross-attention, high accuracy & Slow inference \\ \hline
CUPID & LLM-cleaned text + REP & Best trade-off, interpretable & Requires LLM access \\ \hline
\end{tabular}
}
\label{tab:dbrd_comparison}
\end{table}

\subsection{Fields \& filters}
Our issue tracking dataset provides a diverse set of textual and categorical fields that vary widely in completeness and consistency. We categorize these fields into three buckets based on their nature and usage in our system:

\paragraph{Text for embeddings}
These fields provide the core context of each issue and are used for semantic embedding and similarity computation.
\begin{itemize}
    \item \textbf{Issue Title:} Short textual headline summarizing the problem.
    \item \textbf{Issue Description:} Longer free-text explanation.
    \item \textbf{Root Cause (optional):} Diagnostic explanation of the underlying fault.
    \item \textbf{Feature Summary (optional):} Textual description of the affected subsystem or feature.
\end{itemize}
Issue title and description are always populated. 

\paragraph{Categorical filters}
These fields are consistently populated and serve as the primary dimensions for structured filtering:
\begin{itemize}
    \item \textbf{Program Name:} Identifies the issue domain.
    \item \textbf{Triage Category:} Classification of the issue type.
    \item \textbf{Triage Assignment:} Indicates the team responsible for initial analysis and resolution.
\end{itemize}
These three attributes are highly reliable (populated in nearly all issues) and reflect organizational ownership and functional grouping. Restricting retrieval to issues sharing these categorical attributes ensures that retrieved issues are contextually and organizationally relevant to the query issue.

\paragraph{Other metadata fields}
Several auxiliary fields are inconsistently populated but can provide valuable context when available:
\begin{itemize}
    \item \textbf{Priority, Severity:} Ordinal indicators of impact.
    \item \textbf{Root Cause Category:} Taxonomic classification of defect type.
    \item \textbf{Product Family, Product Name:} High-level identifiers for product lineage.
\end{itemize}

\subsection{Retrieval Strategy and Dual Modes}
\label{embed-summ}

The field variability motivates two retrieval modes that share the same filtering backbone but differ in how textual content is constructed. Across both methods, we compute dense embeddings using \texttt{text-embedding-3-large} \cite{openai_text_embedding_3_large}, stored in a Weaviate vector index \cite{weaviate}. Retrieval is performed via cosine similarity. 

\paragraph{\textit{SIC-Embed} (text-only)}
Efficient, high-throughput retrieval using consistently populated textual fields.
\begin{itemize}
    \item \textbf{Text construction:} \textsc{Title} + \textsc{Description}.
    \item \textbf{Filtering:} query-time filters on \textsc{Program Name}, \textsc{Triage Category}, \textsc{Triage Assignment}.
    \item \textbf{Retrieval:} filtered cosine-similarity search, return top-$k$.
\end{itemize}

\paragraph{\textit{SIC-Summ} (summarization-enhanced)}
Improves semantic coherence by including fields that are sparse or noisy.
\begin{itemize}
    \item \textbf{Text construction:} \textsc{Title} + \textsc{Description} augmented with any populated textual fields \textsc{Root Cause}, \textsc{Feature Summary}, and other metadata.
    \item \textbf{LLM preprocessing:} a large language model (\texttt{o3-mini} \cite{o3-mini}) generates a concise, standardized summary that normalizes terminology and error signatures.
    \item \textbf{Filtering \& retrieval:} same filtered vector search as \textit{SIC-Embed} (top-$k$).
\end{itemize}

\subsection{SIC Pipeline}

Both modes operate within a shared pipeline comprising the following stages:
\begin{enumerate}
    \item \textbf{Preprocessing:} Field extraction, optional LLM-enrichment and normalization.
    \item \textbf{Embedding \& indexing:} Generation and storage of vector representations in Weaviate. We employ an HNSW-based vector index with cosine similarity as the distance metric.
    \item \textbf{Filtering:} Structured pruning using program-level and triage-level metadata.
    \item \textbf{Retrieval:} Cosine similarity search with top-$k$ issue selection. The default value of k is 5.
\end{enumerate}

This design enables low-latency lookups while maintaining flexibility for future model upgrades or additional metadata integration. 

\vspace{-0.1in}
\section{LocAgent-X}
\label{sec:extended-locagent}

\vspace{-0.02in}
\subsection{Baseline LocAgent}
LocAgent \cite{chen-etal-2025-locagent} is a code localization framework that integrates language-aware parsing with retrieval-based indexing. Originally, LocAgent was designed exclusively for Python projects, combining Abstract Syntax Tree (AST) analysis for dependency graph construction and Tree-sitter-based token parsing for BM25 indexing and retrieval. This dual-layer architecture enabled the framework to model both the structural and textual representations of a Python codebase. The AST-based layer captures semantic relationships among program entities such as classes, functions, and imports, while the BM25 index supports efficient token-based retrieval for contextual matching during localization. 

However, industrial codebases are often heterogeneous in which front-end, back-end, and integration layers are implemented in distinct languages. The original Python-only design limited the applicability of LocAgent in such settings. To address this,  LocAgent-X supports additional languages (QML and C++) and further introduces a \emph{mixed-language parsing} mode capable of analyzing multi-language
structures within a single repository. 

Compared to the original LocAgent prompt, we add a small number of language-specific exemplars that demonstrate how to use the existing toolset in C++ and QML. The underlying action space and tool interfaces are unchanged; only the in-context examples differ.

\subsection{Extension Overview}

LocAgent-X required a re-architecting of LocAgent’s parsing pipeline into a language-agnostic abstraction layer. Each supported language implements a dedicated parser interface responsible for extracting comparable semantic constructs (e.g., classes, functions, imports). These language-specific parsers feed into a unified dependency graph builder and a shared BM25 indexing backend, ensuring analysis of  heterogeneous repositories using a consistent representational model.

\subsubsection{Parser Abstraction Layer}
Each language is handled by a specialized parser module that leverages language-appropriate tools internally:

\begin{itemize}
    \item \textbf{Python:} continues to rely on the native AST module to
    construct semantic graphs of function definitions, classes, and imports.
    \item \textbf{C++:} employs the \emph{Tree-sitter} \cite{max_brunsfeld_2025_17180150} C++ grammar to extract
    class hierarchies, namespace relationships, and include dependencies,
    achieving a level of structural depth comparable to AST parsing.
    \item \textbf{QML:} integrates the \emph{Tree-sitter QML/JS} \cite{tree-sitter-qmljs} grammar to
    capture component hierarchies, property bindings, and JavaScript blocks,
    with a lightweight regex-based fallback for resilience in non-standard
    QML constructs.
\end{itemize}

While these parsers use different internal representations and parsing strategies, they all produce NetworkX \cite{networkx} graph structures with standardized node and edge types, enabling the same downstream graph construction and retrieval modules to operate uniformly across languages.

\subsubsection{Unified Dependency Graph Construction}

The dependency graph generator was refactored to operate in one of four modes: \emph{Python-only}, \emph{QML-only}, \emph{C++-only}, or
\emph{mixed-language}. Each language-specific parser emits graph nodes with standardized node types (e.g., \texttt{CLASS}, \texttt{FUNCTION}, \texttt{FILE}), which are subsequently merged into a common graph schema. The graph captures both universal and
language-specific relationships, summarized in
Table~\ref{tab:edges}.

\begin{table}[t]
\centering
\caption{Relationship types in extended dependency graph.}
\label{tab:edges}
\footnotesize
\begin{tabular}{|l|c|c|c|}
\hline
\textbf{Relationship} & \textbf{Python} & \textbf{C++} & \textbf{QML} \\ \hline
Contains & \checkmark & \checkmark & \checkmark \\ \hline
Imports/Includes & \checkmark & \checkmark & \checkmark \\ \hline
Inheritance & \checkmark & \checkmark & hier. \texttt{CONTAINS} \\ \hline
Invocations & \checkmark & \checkmark & -- \\ \hline
\end{tabular}
\end{table}

The \texttt{CONTAINS} relationship operates universally across all languages, modeling the hierarchical containment of functions within classes, or files within directories. In contrast, C++ required specialized handling of edge types to represent its preprocessor-level inclusions and access-modified inheritance relations, which differ semantically from Python’s import and inheritance mechanisms. QML, being declarative, relies primarily on containment and import edges, as its component hierarchies do not follow object-oriented inheritance semantics.

\subsubsection{BM25 Indexing Across Languages}

The BM25 indexing subsystem, originally tailored to Python tokens, was generalized to accept code fragments from any supported language. The Tree-sitter query files define the extraction patterns used for indexable units (such as function definitions, class declarations, or QML component blocks) while preserving syntactic boundaries and semantic integrity. This uniform indexing strategy allows  LocAgent-X to compare and retrieve code snippets across heterogeneous repositories.

\subsubsection{Mixed-Language Parsing and Graph Merging}

To support projects combining multiple languages (e.g., a QML front-end calling C++ back-end logic),  LocAgent-X implements a mixed-language parsing routine.
The process consists of three conceptual stages:

\begin{enumerate}
    \item \textbf{Language detection:} The repository is scanned to identify the
    distribution of source files per language.
    \item \textbf{Parallel parsing:} Each language is parsed independently using
    its corresponding parser module, yielding intermediate graphs.
    \item \textbf{Graph unification:} The individual graphs are merged into a
    single heterogeneous dependency graph through node normalization and edge
    deduplication. Shared entities (e.g., directories or interface files) are
    reconciled by attribute aggregation.
\end{enumerate}

While current cross-language edges are represented implicitly (via shared directories), rather than explicit call or binding edge, the design allows for future extension toward explicit inter-language dependency modeling.

Through these extensions,  LocAgent-X transitions from a Python-specific analyzer to a multi-language code localization framework. The unified design enables graph-based reasoning across multi-language repositories, preserving semantic equivalence while maintaining efficient indexing and querying. This multi-language capability is essential for industrial settings where components developed in Python, QML, and C++ coexist within tightly integrated software ecosystems.

\section{Experimental Evaluation}
\label{sec:evaluation}

For our analysis, we use a large-scale heterogeneous repository (QML, C++) at AMD comprising over 70,000 files and having a complex organizational structure. All evaluation issues are sampled from AMD’s internal tracker and correspond to human-reported defects, as opposed to machine-generated alerts. In contrast, Google’s GITS dataset mixes human- and machine-reported issues, where the latter tend to have \textit{richer} descriptions, and reports substantial performance gaps between these categories \cite{11121734}. Building on this observation, we treat our purely human-reported setting as a harder localization regime: these issues exhibit weaker lexical grounding to the code, which increases localization difficulty.  Below, we describe our experimental evaluation to answer research questions RQ1-RQ3 from Sec.~\ref{sec:introduction}.

\subsection{RQ1: Similar Issue Context}
\label{subsec:rq1}

\textbf{Experimental Setup:} To answer this research question, we collected 2,563 issues from AMD's internal tracker with verified ground truth localizations. 

\paragraph{Rich vs. Sparse} We distinguish between \textit{rich} (verbose, well-structured) and \textit{sparse} (concise, minimally populated) issue descriptions based on the completeness and verbosity of their textual and metadata fields. Rich tickets typically contain detailed descriptions with multiple populated fields, whereas sparse tickets include only a short summary or partial field population. This distinction captures the natural variation in reporting styles within AMD and allows us to evaluate how SIC performance scales with the amount of available contextual information. Our dataset is made of 1,657 rich issues and 906 sparse issues. 

\begin{figure}[t]
\centering
\includegraphics[width=0.99\columnwidth]{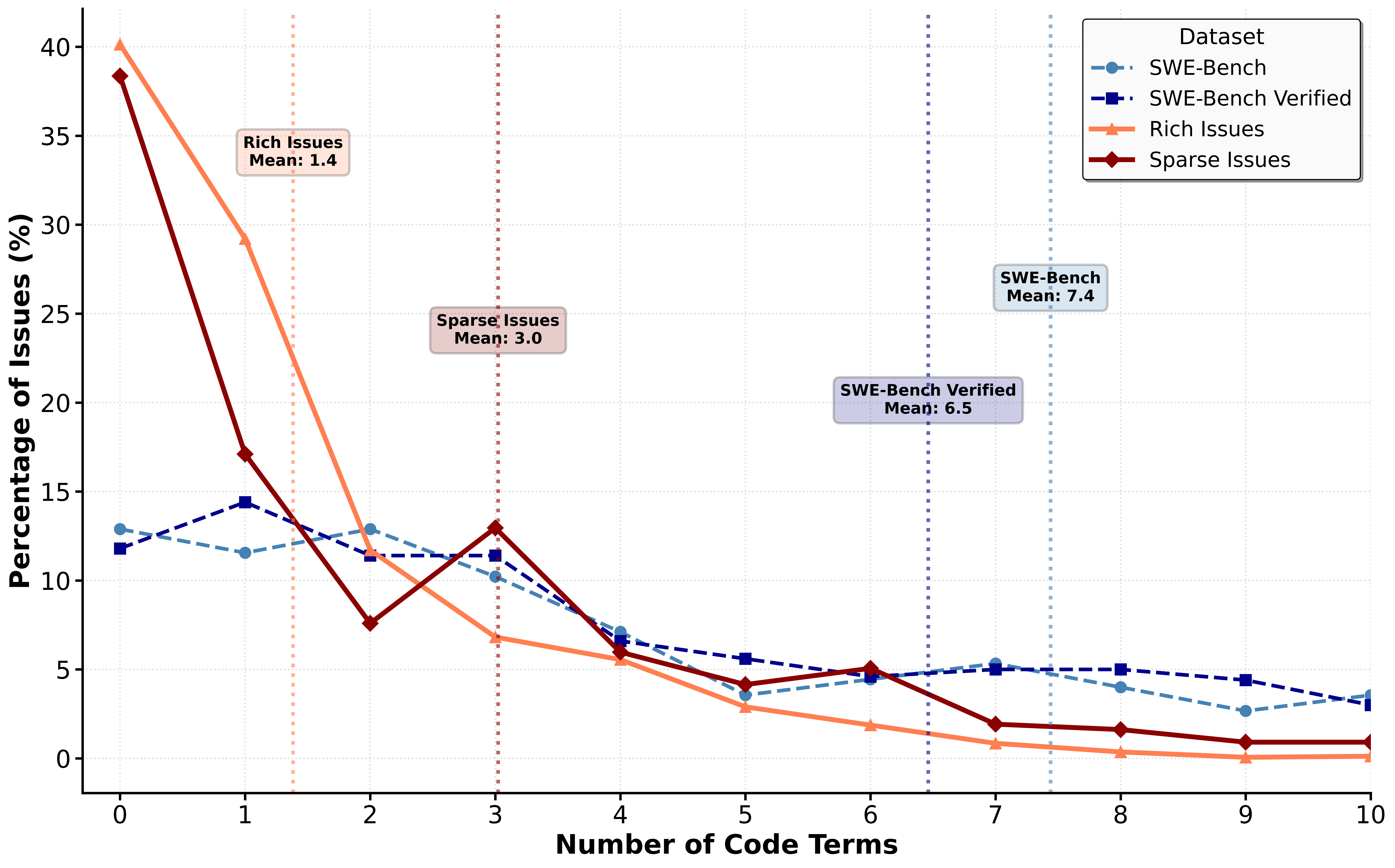}
\vspace{-0.15in}
\caption{Distribution of Code Terms in Issue Descriptions.}
\label{fig:codeterms}
\vspace{-0.1in}
\end{figure}

Consistent with findings from recent work \cite{11121734}, our analysis reveals a fundamental divergence in code term density between industrial and open-source datasets. While 64.2\% of internal issues contain fewer than two code-related terms (snake\_case and CamelCase identifiers), only 24.4\% of SWE-bench and 26.2\% of SWE-bench Verified issues fall into this category. This gap is most pronounced at zero terms: 40.1\% of Rich Issues and 38.4\% of Sparse Issues contain no code terms, three times SWE-bench's rate (12.9\%). The median term count is 1 for internal datasets versus 4 for benchmarks, with only 7.5\% of AMD issues exceeding five terms.

\paragraph{File Path Evaluation Design} We designed a four-level hierarchical similarity evaluation framework to assess how well similar historical issues can guide issue localization. Our approach evaluates similarity at progressively finer granularities, from high-level component matching to exact file identification. Components correspond to individual repositories within the organization, and the top-level directory refers to the first directory under a repository root; all finer-grained hierarchical levels (directory, extension, file) follow the structure illustrated in Figure~\ref{alg:multifile_similarity}.

\begin{figure*}[t]
\begin{tcolorbox}[title=Similarity Computation Algorithm,fonttitle=\bfseries]
\begin{small}
\textbf{Input:} Root issue $R$ with changed files $F_R$, Similar issue $S$ with changed files $F_S$ \\
\textbf{Output:} Similarity score between $R$ and $S$ \\
\textbf{Algorithm:}
\begin{enumerate}[label=\arabic*.]
    \item \textbf{For each root issue and similar issue pair $(R, S)$:}
    \begin{enumerate}[label*=\alph*.]
        \item Split $F_R$ and $F_S$ into file lists
        \item For each root changed file $f_r \in F_R$:
        \begin{enumerate}[label*=\roman*.]
            \item For each similar changed file $f_s \in F_S$, compute:
            \begin{itemize}[leftmargin=1.2em]
            \small
    \item \emph{component\_match} $= \begin{cases}
        1 & \text{if components match} \\
        0 & \text{otherwise}
    \end{cases}$

    \item \emph{top\_dir\_match} $= \begin{cases}
        1 & \text{if top directories match} \\
        0 & \text{otherwise}
    \end{cases}$

    \item \emph{directory\_similarity} $= \dfrac{|\text{matching\_dir\_segments}|}{|\text{total\_dir\_segments}|}$, where \texttt{matching\_dir\_segments} is the number of shared directory segments between $f_r$ and $f_s$, and \texttt{total\_dir\_segments} is the number of directory segments in $f_r$.

    \item \emph{extension\_match} $= \begin{cases}
        1 & \text{if file extensions match} \\
        0 & \text{otherwise}
    \end{cases}$

    \item \emph{exact\_file\_match}\(_{f_r,f_s}\) $= \begin{cases}
        1 & \text{if file names match} \\
        0 & \text{otherwise}
    \end{cases}$

    \item \emph{file\_match} $= I\!\left[\max_{f_r \in F_R,\, f_s \in F_S} \text{exact\_file\_match}_{f_r,f_s} > 0\right]$

    \item \emph{total\_score} $= \text{component\_match} + \text{top\_dir\_match} + \text{directory\_similarity} + \text{extension\_match} + \text{exact\_file\_match}_{f_r,f_s}$
\end{itemize}
            \item Select $f_s^* = \arg\max_{f_s \in F_S} total\_score(f_r, f_s)$ as best match for $f_r$
        \end{enumerate}
        \item Aggregate best match scores for all files in $F_R$
    \end{enumerate}
    \item \textbf{Root Issue Level Aggregation:}
    \begin{enumerate}[label*=\alph*.]
        \item For each $\langle root\_issue, similar\_issue \rangle$ pair, aggregate best match scores for all root files
        \item For each root issue, average $\langle root\_issue, similar\_issue \rangle$ pair similarity scores to produce overall similarity score
    \end{enumerate}
    \item \textbf{Overall Aggregation Across All Root Issues:}
    \begin{enumerate}[label*=\alph*.]
   \item For each $R$, calculate average similarity score across all $S$
   \item Aggregate across all $R$ to produce hierarchical similarity scores
   \end{enumerate}
\end{enumerate}
\end{small}
\end{tcolorbox}
\vspace{-0.1in}
\caption{Algorithm for computing file-level similarity between issue pairs based on hierarchical path structure.
\label{alg:multifile_similarity}}
\vspace{-0.1in}
\end{figure*}

\textbf{Results:}
Table~\ref{tab:similar_issue_performance} demonstrates the effectiveness of hierarchical similarity matching. When ticket descriptions are rich, the results show strong performance at coarse-grained levels (96.02\% component accuracy, 95.70\% top-level directory accuracy) with decreasing but still substantial accuracy at finer granularities (53.14\% file match). As expected, the framework performs more accurately on rich issues, which provide more cues for retrieval. The hierarchical evaluation demonstrates that even when exact file matching is poor, coarse-grained similarity (component and directory level) provides substantial search space pruning. As shown in Table~\ref{tab:rq2-3}, the inclusion of similar-issue context consistently improves localization precision, validating its role as a key intermediate signal in the broader LocAgent framework.

Beyond aggregate similarity, we compare the two retrieval modes introduced in Sec.~\ref{embed-summ}. We find that while the overall performance of \textit{SIC-Summ} and \textit{SIC-Embed} is highly similar, \textit{SIC-Embed} achieves marginally higher localization accuracy, particularly at finer granularities (e.g., File Match for rich issues in Table~\ref{tab:similar_issue_performance}). We attribute this behavior to the prevalence of regression issues in our industrial setting, where a substantial fraction of tickets are repetitions of earlier reports; \textit{SIC-Embed}, with its stronger lexical sensitivity, is better suited to detecting such duplicate issues. Qualitative inspection, however, indicates that \textit{SIC-Summ} can surface more semantically subtle issues that \textit{SIC-Embed} fails to identify. Overall, the two modes are complementary: an exhaustive search benefits from combining both, although at the cost of increased latency and computational overhead introduced by \textit{SIC-Summ}. Therefore, for further experiments, we use \textit{SIC-Embed}.

\begin{table}[t]
\centering
\caption{Hierarchical similarity computation by ticket type.}
\label{tab:similar_issue_performance}
\setlength{\tabcolsep}{3pt}
\small
\begin{tabular}{l cc cc}
\toprule
 & \multicolumn{2}{c}{Rich} & \multicolumn{2}{c}{Sparse} \\
\cmidrule(lr){2-3}\cmidrule(lr){4-5}
Similarity Level & \footnotesize{SIC-Embed} & \footnotesize{SIC-Summ} & \footnotesize{SIC-Embed} & \footnotesize{SIC-Summ}\\
\midrule
Component           & 96.02\% & 95.39\% & 80.78\% & 82.26\% \\
Top-level Directory & 95.70\% & 94.90\% & 48.64\% & 48.64\% \\
Directory Similarity & 67.81\% & 65.88\% & 30.87\% & 28.67\% \\
Extension           & 71.92\% & 70.96\% & 69.10\% & 70.59\% \\
File Match          & 53.14\% & 46.92\% & 37.97\% & 37.20\% \\
\bottomrule
\end{tabular}
\end{table}

\textbf{Takeaway:} Historical organizational context is a valuable signal for constraining the search space and improves localization in both lexical and graph-based pipelines. 

\subsection{RQ2: Multi-Language Generalization}
\label{subsec:rq2}

\textbf{Experimental Setup:}
The evaluation dataset for RQ2 is a subset of the rich issues introduced in Sec.~\ref{subsec:rq1}, we select 140 tickets (67 QML-only issues and 73 C++-only issues). We conducted an ablation study comparing graph-based localization methods under three language configurations (\emph{QML only}, \emph{C++ only}, \emph{mixed} i.e. both QML and C++ only). While the framework is compatible with any language model, all agentic experiments in this study were conducted using Claude 4.5 Sonnet as the backbone LLM.

\textbf{Results:} Table~\ref{tab:rq2-3} reports performance across the three settings. Overall, graph-based methods substantially outperform retrieval-only baselines, indicating that such methods add value for non-Python languages.  Averaged across language configurations on this 140-issue subset, LocAgent-X improves Acc@5 over Code Search by 16.1 points (up to +29.1 in the QML-only setting, 79.10\% vs. 50.00\%).

\textbf{Takeaway:} Graph-based reasoning generalizes  to non-Python code and multi-language repositories, validating LocAgent-X as a portable localization substrate for industrial systems.

\begin{table*}[htbp]
\centering
\caption{Comparison of SIC match rate, Acc@5 and tool-call counts for lexical and graph-based agents}
\label{tab:rq2-3}
\begin{tabular}{l c cc cc cc cc}
\toprule
 &  &
\multicolumn{4}{c}{Lexical Agents} &
\multicolumn{4}{c}{Graph-Based Agents} \\
\cmidrule(lr){3-6}\cmidrule(l){7-10}
\multicolumn{1}{l}{Language} &
\multicolumn{1}{c}{SIC} &
\multicolumn{2}{c}{Code Search} &
\multicolumn{2}{c}{SIC + Code Search} &
\multicolumn{2}{c}{LocAgent-X} &
\multicolumn{2}{c}{Multi-CoLoR} \\
\cmidrule(lr){2-2}\cmidrule(lr){3-4}\cmidrule(lr){5-6}\cmidrule(l){7-8}\cmidrule(l){9-10}
 & Match Rate
 & Acc@5 & \# Tool Calls
 & Acc@5 & \# Tool Calls
 & Acc@5 & \# Tool Calls
 & Acc@5 & \# Tool Calls \\
\midrule
QML only
 & 43.47\%
 & 50.00\% & 20
 & 56.06\% & 15
 & 79.10\% & 21
 & \textbf{83.58\%} & 17 \\
C++ only
 & 64.55\%
 & 44.28\% & 18
 & 55.71\% & 17
 & 57.53\% & 23
 & \textbf{75.34\%} & 18 \\
QML and C++
 & 55.10\%
 & 56.08\% & 20
 & 58.10\% & 17
 & 62.14\% & 22
 & \textbf{63.33\%} & 21 \\
\bottomrule
\end{tabular}
\end{table*}

\subsection{RQ3: Combined Impact Analysis}
\label{subsec:rq3}

\textbf{Experimental Setup:} We used the same dataset as in answering RQ2. Through a comprehensive ablation across all component combinations (SIC, Code Search, SIC+Code Search, LocAgent-X, and Multi-CoLoR), we aimed to isolate the contribution of organizational context and graph-based reasoning, as well as their combined impact. Code Search is a lexical baseline where the agent has access only to a standard code search and file-viewing (via GitHub tools) over the repository. 

We distinguish between how language constraints are applied for graph-based (LocAgent-X, Multi-CoLoR) and lexical (Code Search, SIC+Code Search) agents. For the graph-based methods, we explicitly control the indexed file set. In the \emph{QML only} condition, we build the structural code graph over only \texttt{.qml} files under the repository root; in the \emph{C++ only} condition, the index is restricted to \texttt{.cpp} (and associated header) files. In the \emph{mixed} condition, LocAgent-X and Multi-CoLoR index the full \texttt{.qml}+\texttt{.cpp} subtree, yielding a unified multi-language graph. The lexical methods operate over the same repository but are constrained via natural-language prompts. In the \emph{QML only} and \emph{C++ only} settings, we add an explicit instruction constraining the search to the appropriate language slice. In the mixed-language setting, we do \emph{not} specify any target language in the prompt.

\textit{Match Rate} (\emph{file\_match} in Fig.~\ref{alg:multifile_similarity}) and \textit{Acc@5} both measure whether the predicted file set intersects the ground-truth fix. Specifically, they are defined as the fraction of evaluation issues for which at least one of the 5 retrieved files overlaps with the ground-truth. For each configuration, we additionally report \textit{\# Tool Calls}, defined as the total number of invocations of any code-localization tool available to the agent (each call to the SIC retriever, any tool in the LocAgent-X toolset, or GitHub Code Search).

\textbf{Results:}
Acc@5 trends follow a clear hierarchy (Table~\ref{tab:rq2-3}):
\textbf{(1)} SIC alone is competitive with, and for C++-only issues slightly better than, baseline Code Search, confirming that historical anchors are useful even without explicit code reasoning.
\textbf{(2)} Incorporating SIC consistently enhances both lexical and graph-based pipelines while reducing tool calls.
\textbf{(3)} LocAgent-X outperforms both lexical agents in every subset, highlighting the benefit of dependency-graph traversal.
\textbf{(4)} These gains are language-sensitive: SIC contributes most on C++ issues, while graph-based reasoning contributes most on QML; for heterogeneous issues, strong performance requires both historical and graph-based signals.
\textbf{(5)} Multi-CoLoR achieves the best overall performance across all settings.

\textbf{Takeaway:}
The combination of organizational context retrieval and graph-based reasoning yields the most reliable localization results. SIC narrows the search space, while LocAgent-X provides structured exploration, and their integration in Multi-CoLoR delivers robust performance in  heterogeneous industrial repositories.

\subsection{Threats to Validity}
\label{limits}

Despite promising results, several factors may influence the generality of our findings.

\textbf{Internal validity.}
Ground-truth localizations are derived from verified commits and may contain minor inconsistencies (e.g., additional non-essential files modified in the same change). In addition, the behavior of \textit{SIC-Embed} and \textit{SIC-Summ} is coupled to specific design choices; alternative summarization strategies, embedding models, or index configurations could therefore shift the observed relative gains, even on the same dataset.

\textbf{External validity.}
Although we generalize graph construction to C++ and QML, our experiments are conducted within a single organization, on a specific codebase, with a fixed language mix. Results may not transfer directly to ecosystems with different issue-tracking conventions, repository structures, programming languages, or inter-language binding mechanisms. Replication on public multi-language benchmarks (e.g., Multi-SWE-bench) and across diverse industrial contexts is needed to confirm cross-domain generalization.

\textbf{Construct validity.}
Our evaluation metrics, hierarchical path similarity and \textit{Acc@k}, capture localization accuracy but not developer effort or interpretability. Future studies incorporating human-in-the-loop assessment or time-to-resolution measures could provide more practical insights.

\section{Lessons Learned}
\label{sec:discussion}
Our evaluation yields several actionable insights that extend beyond this specific implementation. While Multi-CoLoR provides one instantiation, the underlying principles generalize to other organizations and tooling ecosystems. These lessons highlight what worked and what appears generalizable beyond AMD to similar industrial contexts.

{\bf Organizational context is a first-class localization signal.}  Through answering RQ1, we find that incorporating historical organizational knowledge substantially improves localization performance: SIC consistently narrows the search space and boosts accuracy for both lexical and graph-based methods. These gains arise because industrial issue trackers encode stable, organization-specific ownership and subsystem patterns that are not visible from issue descriptions alone.

Importantly, \emph{SIC is only one mechanism for retrieving organizational knowledge}. Other organizations may instead rely on commit-history clustering, test-failure signatures, authorship patterns, defect taxonomies, or project-specific documentation. The general insight is that effective localization systems should incorporate some form of historical memory, regardless of its specific implementation. The key requirement is that this component must transform noisy, semi-structured issue descriptions into actionable contextual signals that meaningfully constrain the downstream search space.

{\bf Graph-based reasoning generalizes and reduces reliance on issue text alone.}
Answering RQ2 showed that graph-based reasoning generalizes beyond Python and remains effective even when issue descriptions contain few or no code identifiers. In our implementation, we follow LocAgent’s design by using lightweight, language-specific parsers to construct a unified dependency graph capturing relationships across QML and C++ codebases, paired with BM25-indexed search and retrieval over code tokens to support entity lookup and extraction.
Alternative implementations may employ different graph-construction strategies (e.g., AST-level graphs, call/reference graphs, summary graphs) or different language parsers. The essential requirement is the ability to navigate code relationships that are not apparent from textual similarity alone. Likewise, the search component need only enable efficient retrieval of entities and their contents; this module has the lowest implementation complexity and can be readily substituted with existing tools such as GitHub Code Search \cite{github_code_search} or Sourcegraph \cite{sourcegraph_about}.

In summary, this component must support (i) traversal over meaningful structural relationships in the repository graph and (ii) reliable entity-level lookup and retrieval of graph contents, regardless of the specific parser, graph representation, or search backend used.

{\bf Composing organizational and structural signals yields the most reliable localization.}
Answering RQ3 showed that organizational context and graph-based reasoning contribute \emph{complementary} gains. Neither historical context nor structural reasoning is sufficient in isolation; the most reliable localization arises when they are composed.

This leads to a broader design principle: localization systems should treat context retrieval, structural reasoning, and search as distinct modules that can be composed, rather than as a single monolithic model. In our instantiation, these roles are filled by SIC, LocAgent-inspired graph traversal and BM25-based code search, but each can be instantiated differently depending on organizational constraints.

The modular design enables flexible adoption in industrial settings. Because each component exposes a narrow, well-defined interface (e.g., passing a pruned set of files, ranked candidates, or graph nodes), organizations can substitute implementations to fit their environment (for example, replacing SIC with commit-history–based retrieval when issue metadata is weak, or using proprietary code search backends instead of BM25 indices). The same structure supports incremental deployment: teams may start with code search alone, then add graph-based traversal, and later integrate historical context retrieval as organizational data and infrastructure mature. Finally, the plug-and-play architecture naturally accommodates additional signals such as test failures, runtime profiling, documentation, or developer annotations, allowing multi-modal evidence to be incorporated without redesigning the pipeline.

{\bf Scalability and operational constraints shape deployment strategy.}
Our experiments highlight that scalability and latency considerations strongly influence how localization systems should be deployed in practice. Building and maintaining historical indexes (for SIC) and repository graphs (for graph-based reasoning) introduces non-trivial preprocessing and storage costs. While these costs are acceptable for large, relatively stable codebases where indexes can be reused across many issues, they are less suitable for highly volatile components or tight CI/CD loops where frequent re-indexing would dominate latency.
These observations suggest that organizations should adopt context-heavy components, such as historical issue retrieval, selectively and prioritize long-lived, stable code regions for full indexing.

{\bf Practical deployments benefit from clear use cases.}
Finally, our experience indicates that practical adoption is easier when the localization system is aligned with concrete use cases. The same core architecture supports at least three deployment modes: (i) as a standalone localization tool that accepts issue reports and returns ranked file locations to assist developers during triage; (ii) as a preprocessing stage within APR pipelines, where narrowing the search space directly benefits downstream repair; and (iii) as part of hybrid human-AI workflows, where developers interactively refine suggestions and inspect reasoning paths provided by the graph-based traversal. These scenarios reflect how such systems are most naturally integrated into existing industrial workflows.

\section{Conclusion and Future Work}
\label{sec:conclusion}

This work presented Multi-CoLoR, a unified framework for context-aware code localization across large, multi-language repositories. By coupling SIC with LocAgent-X, Multi-CoLoR bridges organizational knowledge and graph-based reasoning. On a large industrial codebase, SIC alone provides strong coarse-grained guidance and search-space pruning, graph-based reasoning generalizes effectively beyond Python, and their combination yields the highest Acc@5 with fewer tool calls. The framework is deliberately modular: context retrieval, graph traversal, and lexical code search are exposed as interchangeable components that can be instantiated with different backends and integrated into existing industrial workflows.

While Multi-CoLoR demonstrates strong localization performance on industrial codebases, several opportunities remain to broaden its generality and further improve efficiency:

\textbf{Language and Parser Generalization.}
A key direction for future work is improving generalizability across languages and ecosystems. Although SIC is language-agnostic at the text level, LocAgent-X relies on language-specific parsers to construct the unified dependency graph. One avenue is to explore more universal parsing abstractions that reduce per-language engineering effort. In the near term, extending coverage to the broader set of Tree-sitter-supported languages would already make Multi-CoLoR applicable to domains such as mobile (Java/Kotlin/Swift), back-end (Go/Rust), and modern web stacks (TypeScript/JavaScript), and enable comparative evaluation on multilingual public benchmarks.

\textbf{Improving Similar-Issue Summarization.}
Our results indicate that summarization-enhanced retrieval (\textit{SIC-Summ}) is competitive but data-sensitive: lexical embeddings perform better on repeated regression reports, whereas summaries have the potential to better capture subtle or semantically shifted issues. Future work will explore richer summarization strategies and model families, including structured summaries that explicitly encode suspected components, regression signatures, or error patterns. An interesting direction is to jointly optimize the summarization and retrieval stages, learning summaries that preserve the lexical cues needed to catch regressions while still abstracting away noise. Adaptive or hybrid schemes that combine \textit{SIC-Embed} and \textit{SIC-Summ}, for example, by routing issues based on sparsity, field completeness, or historical performance, may yield more robust behavior across heterogeneous issue distributions.

\textbf{Integration into End-to-End Repair Pipelines.}
Finally, we plan to integrate Multi-CoLoR into complete APR and continuous-integration pipelines. In such settings, localization quality is only one factor; developer trust, interaction patterns, and downstream patch success all matter. Embedding Multi-CoLoR as a pluggable localization stage in existing agentic repair systems would allow us to quantify how improvements in Acc@k translate into higher end-to-end resolution rates and lower developer effort. This integration also opens opportunities to incorporate additional signals  into both SIC and graph traversal, enabling the system to refine its organizational memory and structural priors over time.

\begingroup
\renewcommand\thefootnote{}
\footnotetext{AMD, the AMD Arrow logo and combinations thereof are trademarks of Advanced Micro Devices, Inc.}
\endgroup

\bibliographystyle{ieeetr} 
\bibliography{refs} 

\end{document}